\documentclass[11pt]{article}
\usepackage[english]{babel}
\usepackage{amssymb,amsfonts,amsmath}

\newcommand{\ff}{\mathcal{F}(G,\mathbf{C})}

\begin{document}

\title{Remarks on the star product of functions on finite and compact groups}

\author{P. Aniello$^{1,2}$, A. Ibort$^3$, V.I. Man'ko$^4$, and G. Marmo$^1$
\\
{\small $^1$ Dipartimento di Scienze Fisiche
dell'Universit\`a di Napoli `Federico II'}\\  {\small and Istituto Nazionale di
Fisica Nucleare (INFN) -- Sezione di Napoli,}\\ {\small Complesso
Universitario di Monte S.\ Angelo, via Cintia, 80126 Napoli, Italy}
\\ {\small  $^2$ Facolt\`a di Scienze
Biotecnologiche, Universit\`a di Napoli `Federico II', Napoli, Italy
} \\ {\small $^3$ Depto.\ de Matem\'aticas, Univ.\ Carlos III de
Madrid, 28911 Legan\'es, Madrid, Spain} \\{\small $^4$ P.N. Lebedev
Physical Institute, Leninskii Prospect 53, Moscow 119991, Russia}}

\maketitle

\begin{abstract}
Using the formalism of quantizers and dequantizers, we show that the
characters  of irreducible unitary representations of finite and
compact groups provide kernels for star products of complex-valued
functions of the group elements. Examples of permutation groups of
two and three elements, as well as the $SU(2)$ group, are
considered. The $k$-deformed star products of functions on finite
and compact groups are presented. The explicit form of the
quantizers and dequantizers, and the duality symmetry of the
considered star products are discussed.
\end{abstract}

PACS: 03.65.-w; 03.65.Wj

Keywords: star product, finite groups, irreducible representations,
characters.


\section{Introduction}

Traditionally the finite symmetry groups and their irreducible
representations are used to describe the properties of crystals
and electrons in solids. Also for the description of phase
transitions, one needs to know the change of symmetry structure
and corresponding group representation properties. In the last
years, there is a growing interest in constructing quantum
mechanics of finite or discrete phase spaces
\cite{Schwin,Beppe,Vourdas}. Besides, in the context of quantum
computations \cite{Nielsen} one often considers finite-dimensional
Hilbert spaces associated with qubit or, more generally, qudit
states. Therefore, it is quite natural to study the phase-space
realizations of quantum systems suitably associated with
representations of finite or compact groups. The description of
states of a quantum system by means of Wigner
(quasi-)distributions in the case of continuous position and
momentum variables has been considered in a wide variety of
contexts \cite{Wigner32,Husimi,Sudar,Glauber}. It relies on
considering the phase space (a symplectic vector space) as a
quotient of the Heisenberg--Weyl group by its centre. Thus a
unitary representation of this group --- or, equivalently, a
projective unitary representation of the vector group  (the
symplectic vector space) --- may be regarded as an immersion of
this space as a smooth submanifold of the Weyl algebra generated
by the unitary operators. This particular immersion allows to
pull-back the $\mathbf{C}^*$-algebra of operators and, therefore,
allows to induce a star product on the functions on the phase
space (see, e.g.,~\cite{Strat,Zachos,Berezin}). This product may
be expressed by means of a kernel function constructed out of the
unitary operators associated with points of the symplectic vector
space. This procedure may be generalized to any manifold as long
as a suitable orthonormality condition is implemented. We have
considered several instances of this procedure in the
past~\cite{Simoni,OlgaJPA,Patri1,Patri2}. The Weyl--Wigner
approach we have described has been considered very often also for
quantum systems with a finite-dimensional carrier Hilbert
space~\cite{Schwin,Beppe,Vourdas}.

The aim of this paper is to find the connection of the properties
of quantum systems, which have symmetries described by finite or
compact symmetry groups (crystallographic groups, permutation
groups, rotation group), with the star-product quantization
approach. In the context of mathematical formulation, the aim of
this paper is to consider the immersion of finite or compact
groups in the space of unitary operators acting on some Hilbert
space and pull-back the $\mathbf{C}^*$-algebra of operators to
describe nonlocal and noncommutative products on the space of
functions. Such mathematical construction provides the possibility
to discuss the properties of quantum systems associated with
physical observables identified with operators using a
classical-like approach where the observables are identified with
functions on phase space. But the multiplication rule for these
functions is determined by a specific star-product procedure. Some
examples of crystallographic and permutation groups are presented
to illustrate the procedure.

The paper is organized as follows.

In Sect.~{\ref{general}}, we recall some basic facts about star
products. In Sect.~{\ref{finite}}, we focus on the special case of
finite groups. Next, in Sect.~{\ref{formulae}}, some formulae
involving characters are derived, and we illustrate our results by
means of examples in Sect.~{\ref{examples}}. Finally, in
Sect.~{\ref{conclusions}}, conclusions are drawn.


\section{General aspects of star products}\label{general}

The construction of a Weyl system, when considered from the point of view of the immersion
of the symplectic vector space into the group of unitary operators ${\cal U}({\cal H})$ acting in some Hilbert
space ${\cal H}$, may be described in the following way.
We consider a manifold $M$ and a couple of maps $~U\colon M\to {\cal U}({\cal H})~$ and
$~D\colon M\to {\cal U}({\cal H})~$ usually called
\emph{dequantizer} and \emph{quantizer}, respectively, with the following property:
\begin{equation}\label{1}
\mbox{Tr}\,\hat U(\vec x)\hat D(\vec x')=\delta(\vec x-\vec x').
\end{equation}
With any operator $\hat A$ acting in the Hilbert space ${\cal H}$
we can associate a function on $M$ by setting:
\begin{equation}\label{2}
f_A(\vec x)=\mbox{Tr}\left(\hat A\hat U(\vec x)\right).
\end{equation}
Conversely, with each function one associates an operator by
setting:
\begin{equation}\label{3}
\hat A=\int f_A(\vec x)\hat D(\vec x)\,d\vec x.
\end{equation}
The role the operator-valued maps $\hat U$ and $\hat D$ (that we will also call `basic operators' in the following)
play in these formulae explains
their names.

The star product of functions induced by the operator product is defined by
\begin{equation}\label{4}
\left(f_A\star f_B\right)(\vec x):=f_{AB}(\vec x).
\end{equation}
The kernel function or `structure constants' implementing the
associative product has the following expression involving the
couple quantizer--dequantizer:
\begin{equation}\label{5}
K(\vec x,\vec y,\vec z)=\mbox{Tr}\left(\hat D(\vec y)\hat D(\vec
z)\hat U(\vec x)\right).
\end{equation}
hence:
\begin{equation}\label{5-bis}
\left(f_A\star f_B\right)(\vec x)=\iint K(\vec x,\vec y,\vec z) f_A(\vec y) f_B(\vec z)\,d \vec y d\vec z.
\end{equation}
From the definition we find that the associativity condition is trivially
satisfied:
\begin{equation}\label{6}
\left(f_A\star f_B\right)\star f_C=f_{A}\star\left(f_B\star f_C\right).
\end{equation}

At this point, the skew-symmetrization provides us, in a natural way, with a Lie algebra of
functions on $M$ defined by means of the integral kernel
\begin{equation}\label{7}
C\left(\vec x,\vec y,\vec z\right)=\mbox{Tr}\left(\left[\hat D(\vec y),\hat
D(\vec z)\right]\hat U(\vec x)\right),
\end{equation}
which, in a  synthetic way, may be written as
\begin{equation}\label{8}
C\left(\vec x,\vec y,\vec z\right)\rightarrow C^{\vec x}_{\vec
y\vec z}.\end{equation} It should be noticed that, restricting to
real part of the algebra, the symmetrized product provides us with
a Jordan algebra. The compatibility condition between the two
products would then provide us with a Lie--Jordan algebra.


\section{The case of finite groups} \label{finite}

Let us restrict now our attention to finite groups. Let $G$ be a
group with $N$ elements, $G=\left\{g_1,g_2,\ldots,g_N\right\}$. It
is well known that all the irreducible representations of such a
group are finite-dimensional and unitarizable and satisfy the
orthogonality conditions (see, for example \cite{Clemente,Weyl})
\begin{equation}\label{9}
\sum_{k=1}^{N}u^{(s)}_{mn}(g_k)u^{*(p)}_{\alpha\beta}(g_k)=\delta_{m\alpha}
\delta_{n\beta}\frac{N}{N_s}\delta_{sp},
\end{equation}
where $N_s$ is the dimension of the representation $u^{(s)}$ (the dimension of the vector space where
$u^{(s)}$ acts).
We may replace previous association $\vec x\to\hat U(\vec x)$,
$\vec x\to\hat D(\vec x)$ with maps $G\to {\cal U}({\cal H})$ given by
\begin{equation}\label{10}
g_k\to u(g_k),\qquad g_k\to u^{-1}(g_k)\frac{N}{N_s}\,.
\end{equation}
With the help of these maps, we can define complex-valued
functions on $G$, forming the group algebra (see, e.g.,
\cite{Clemente,Weyl}), associated with operators (matrices) on
${\cal H}^{(s)}$ by setting
\begin{equation}\label{11}
f_A^{(s)}(g_k)=\mbox{Tr}\left(Au^{(s)}(g_k)\right)=\sum_{m=1}^{N_s}\left(A
u^{(s)}(g_k)\right)_{mm}
\end{equation}
or
\begin{equation}\label{12}
f_A^{(s)}(g_k)=\sum_{m,j=1}A_{mj}u^{(s)}(g_k)_{jm}.
\end{equation}
Again the ``reconstruction'' of the matrix from the function is provided by
\begin{eqnarray}\label{13}
A_{nj}&=&\frac{N_s}{N}\sum_{k=1}^N f_A^{(s)}(g_k)u^*_{jn}(g_k)\nonumber\\
&=&\frac{N_s}{N}\sum_{k=1}^N f_A^{(s)}(g_k)u^{-1}_{nj}(g_k)\nonumber\\
&=&\frac{N_s}{N}\sum_{k=1}^N f_A^{(s)}(g_k)u_{nj}(g_k^{-1}).
\end{eqnarray}
This shows that there is a one-to-one correspondence between complex-valued functions on $G$
and operators in $\mathcal{H}^{(s)}$. We should stress that the operator associated with a given function
depends on the chosen representation.

The kernel of the star product corresponding to the operator product is given by
\begin{equation}\label{14}
K(g_1,g_2,g_3)=\mbox{Tr}\left\{
\left(\frac{N_s}{N}\right)^2u^{-1}(g_2)u^{-1}(g_3)u(g_1)\right\}.
\end{equation}
Recalling the definition of characters of a representation, we find that up to normalization
the kernel function is represented by a character of the group $G$ that we are considering.

The associative algebra generated by these basic operators has
structure constants given by
\begin{equation}\label{15}
a_{as}^c\equiv\frac{\mbox{Tr}\left(u(g_a)u(g_s)u^{-1}(g_c)\right)}{\mbox{Tr}\,{\mathbf
1}_k}\,.\end{equation}
 If we use a deformed product \cite{OlgaJPA} by
means of a fixed unitary transformation ${k}$, we have
\begin{equation}\label{16}
a_{as}^c(k)\equiv\frac{\mbox{Tr}\left(u(g_a)ku(g_s)ku^{-1}(g_c)\right)}{\mbox{Tr}\,{\mathbf
1}_k}\,.\end{equation}

Following \cite{Patri2} we can also define the dual star product by exchanging the role
of quantizers and dequantizers
\begin{equation}\label{18}
\hat U_d(\vec x)\rightarrow\frac{N_s}{N}\,u^{-1}(g_k)
\end{equation}
and
\begin{equation}\label{19}
\hat D_d(\vec x)\rightarrow u(g_k).
\end{equation}

Thus, our main result amounts to say that structure constants of
the associative product induced on finite groups by operators
acting on some Hilbert space carrying an irreducible
representation are given by characters of the group representation
we are considering.

Therefore, the tables of characters available in the literature
allows us to construct in explicit manner families of associative
products on ${\cal F}(G,{\mathbf C})$.

The construction we have considered shows, very clearly that it is not necessary that
the map $G\to{\cal U}({\cal H})$ be a group representation. Indeed we may consider a set
$S$ with a measure $ds$ and an algebra ${\cal A}$ of operators and require that
$$
\mbox{Tr}\,\hat D(s)\hat U(s^\prime)=\delta(s,s^\prime),$$
where $\delta(s,s^\prime)$ stays for a Kronecker $\delta~$ or a Dirac delta as the case may be.

Then out of the basic operators we may define
$$f_A(s)=\mbox{Tr}\left(A\hat U(s)\right)$$
along with $$\left(f_a\star f_b\right)(s)=\mbox{Tr}\,\hat A\hat B\hat U(s).$$
The reconstruction of $\hat A$ is again permitted by using $\hat D(s)$, it is
$$\hat A=\int f_A(s)\hat D(s)\,ds.$$
Of course, when $S$ is finite-dimensional the measure will be a concentrated measure
and the integral is replaced by a sum.

Elsewhere \cite{OlgaJPA} we have considered deformations of the
operator product by setting
$$\hat A\cdot_{\hat{K}}\hat B=\hat A\cdot\hat K\cdot\hat B.$$
If we use this deformed product on the space of operators, we
induce a deformed product also on the associated functions on $S$
as given in (\ref{16}).

These observations should be kept in mind when we want to classify
associative products on $\ff$. We should stress, however, that the
identification of the structure constants with characters requires
that we consider a group $G$ and its unitary representations.


\section{Some formulae for characters of finite groups} \label{formulae}

It is now possible to use the identification of structure constants with characters
 to derive easily some identities that characters must satisfy. The function associated with the unity
operator will be just the character of irreducible representation
\begin{equation}\label{22}
f_{\mathbf I}(g_k)=\mbox{Tr}\left({\mathbf I}u(g_k)\right)=\chi(g_k).
\end{equation}
The relation $~{\mathbf I}\cdot{\mathbf I}={\mathbf I}~$ implies
\begin{equation}\label{24}
\frac{N_s^2}{N^2}\sum_{k,s=1}^N\chi(g_k)\chi(g_s)
\chi\left(g_3g_k^{-1}g_s^{-1}\right)=\chi(g_3)
\end{equation}
and similarly for the dual star product.

Thus one has identity (\ref{24}) which must be satisfied by
characters of irreducible representations. On the other hand, the
dual star-product scheme yields
\begin{equation}\label{25}
\left(\frac{N_s}{N}\right)^2\sum_{k,k'=1}^N\chi(g_k^{-1})\chi(g_{k'}^{-1})
\chi\left(g_kg_{k'}g_{s}^{-1}\right)=\chi(g_s^{-1}).
\end{equation}
Since
\begin{equation}\label{26}
{\mathbf 1}\cdot g=g,
\end{equation}
one has
\begin{equation}\label{27}
\frac{N_s^2}{N^2}\sum_{g_kg_{k'}=1}^N\chi(g_k)\chi(gg_{k'})
\chi\left(g_sg_k^{-1}g_{k'}^{-1}\right)=\chi(gg_s).
\end{equation}
Another composition formula for characters of finite (or compact)
groups
\begin{equation}\label{weyl}
\chi(g_s)\chi(g_t)=\frac{N}{N_s}\sum_r\chi\left(g_sg_r^{-1}g_tg_r\right)
\end{equation}
is presented, for example, in \cite{Clemente,Weyl}. One can see
that our formulae (\ref{25}) and (\ref{27}) are consistent with
(\ref{weyl}).


\section{Examples} \label{examples}

\subsection{A group with two elements}

Let us consider the reflection group $G=\left\{{\mathbf I},P\right\}$ containing the
identity and the reflection.

The group can be realized as group of two matrices
$$
g_1=\left(\begin{array}{cc}
1&0\\0&1
\end{array}\right),\qquad
g_2=\left(\begin{array}{cc}
0&1\\1&0
\end{array}\right).$$
The space of functions
$$f:\left\{{\mathbf I},P\right\}\to{\mathbf C}$$
is isomorphic to ${\mathbf C}^2$, therefore the associative products on these functions
can be considered as products on vectors of a two-dimensional complex vector space. If we use Dirac notation, we find
\begin{equation}\label{R5}
\mid f\rangle=\left(\begin{array}{c} f(1)\\f(2)\end{array}\right).
\end{equation}
The product of two functions $f_1,f_2$ is given by
\begin{equation}\label{R5a}
\left(f_1\star f_2\right)(k)=\sum_{k_a,k_b=1}^2
K(k_a,k_b,k)f_1(k_a)f_2(k_b),
\end{equation}
the kernel function (or the structure constants) is
\begin{equation}\label{R7}
K(k_a,k_b,k)=\chi\left(g_ag_bg_k^{-1}\right).\end{equation}

The reflection group $G$ contains two elements ---
identity $I=g_1$ and reflection $P=g_2$. There are two irreducible
one-dimensional representations $R^{(s)}$ given in table
\begin{equation}\label{R1}
\begin{array}{cccc} g_1&g_2\\1&1\end{array}
\end{equation}
and
\begin{equation}\label{R2}
\begin{array}{cccc} g_1&g_2\\1&-1\end{array}.
\end{equation}

Thus this group can create two-dimensional Lie algebra. Also this
group can be considered as permutation group of two elements
\begin{equation}\label{R3}
g_1=12,\qquad g_2=21.
\end{equation}
Any function on the group $G$ has two values
\begin{equation}\label{R4}
f(1)\equiv f(g_1),\qquad f(2)\equiv f(g_2).
\end{equation}

The characters in (\ref{R1}) and (\ref{R2}) coincide with matrix
elements. So our kernel being a function of three variables each
having two values reads for representation (\ref{R1})
\begin{equation}\label{R8}
K(g_{k_1},g_{k_2},g_k)=1.\end{equation} Thus in this case, one has
\begin{equation}\label{R9}
f_1(k)\star
f_2(k)=f_1(g_1)f_2(g_1)+f_1(g_1)f_2(g_2)+f_1(g_2)f_2(g_1)+f_1(g_2)f_2(g_2).
\end{equation}
In the vector form, we get the result of star-product of two
vectors
\begin{equation}\label{R10}
\mid\psi\rangle_1\star\mid\psi\rangle_2=\mid\psi\rangle,
\end{equation}
where vector $\mid\psi\rangle$ has equal components
\begin{equation}\label{R11}
\mid \psi\rangle=\left(\begin{array}{cc} f\\f\end{array}\right)
\end{equation}
and
\begin{equation}\label{R12}
f=f_1(1)f_2(1)+f_1(1)f_2(2)+f_2(1)f_1(2)+f_2(1)f_1(2).
\end{equation}

For the case of representation (\ref{R2}), one has the kernel
\begin{equation}\label{R13}
K(g_1,g_1,g_1)=1,\quad K(g_1,g_1,g_2)=-1,\quad
K(g_1,g_2,g_1)=-1,\quad K(g_1,g_2,g_2)=1.
\end{equation}
This kernel provides the result of the product
\begin{equation}\label{R14}
\mid \tilde \psi\rangle=\mid \tilde \psi_1\rangle\star\mid \tilde
\psi_2\rangle,
\end{equation}
where vector $\mid\tilde\psi\rangle$ has two components
\begin{equation}\label{R15}
\mid \tilde \psi\rangle=\left(\begin{array}{c} \tilde f\\-\tilde f
\end{array}\right),
\end{equation}
where
\begin{equation}\label{R16}
\tilde f=
f_1(1)f_2(1)-f_1(1)f_2(2)-f_1(2)f_2(1)+f_1(2)f_2(2).
\end{equation}
The structure constants of Lie algebra obtained by means of the
kernels of the associative products are equal to zero for both
kernels
\begin{equation}\label{R17}
C_{\alpha\beta}^\gamma=0.
\end{equation}
Thus we got Abelian algebras of dimension two.

We give explicit forms of these products, point-wise product, and
standard convolution product.

For example, point-wise product of two vectors
$\left(\begin{array}{c}x_1\\x_2
\end{array}\right)$ and $\left(\begin{array}{c}y_1\\y_2
\end{array}\right)$
gives
\begin{equation}\label{R18}
\left(\begin{array}{c}x_1\\x_2
\end{array}\right)\star\left(\begin{array}{c}y_1\\y_2
\end{array}\right)=
\left(\begin{array}{c}x_1y_1\\x_2y_2
\end{array}\right)\end{equation}
and convolution product gives
\begin{equation}\label{R19}
\left(\begin{array}{c}x_1\\x_2
\end{array}\right)\star\left(\begin{array}{c}y_1\\y_2
\end{array}\right)=
\left(\begin{array}{c}x_1y_1+x_2y_2\\x_2y_1+x_1y_2
\end{array}\right).\end{equation}
The constructed products give correspondingly
\begin{equation}\label{R18a}
\left(\begin{array}{c}x_1\\x_2
\end{array}\right)\star\left(\begin{array}{c}y_1\\y_2
\end{array}\right)=
\left(\begin{array}{c}x_1y_1+x_1y_2+x_2y_1+x_2y_2\\
x_1y_1+x_1y_2+x_2y_1+x_2y_2
\end{array}\right)\end{equation}
and
\begin{equation}\label{R19a}
\left(\begin{array}{c}x_1\\x_2
\end{array}\right)\star\left(\begin{array}{c}y_1\\y_2
\end{array}\right)=
\left(\begin{array}{c}
x_1y_1-x_1y_2-x_2y_1+x_2y_2\\
-x_1y_1+x_1y_2+x_2y_1-x_2y_2
\end{array}\right).\end{equation}

One can see that product (\ref{R19}) is compatible\footnote{We
recall that two associative products are said to be
\emph{compatible} if a linear combination of their structure
constants defines again an associative product \cite{Grabowski}.}
with the convolution product (\ref{R18a}) provided that the
vectors satisfy the condition: $~x_1=x_2~$ and $~y_1=y_2$.
Analogously, for vectors under the condition $~x_1=-x_2~$ and
$~y_1=-y_2$ product (\ref{R18a}) is compatible with (\ref{R19}).
This means that the two products (\ref{R19}) and (\ref{R19a}) are
compatible. One can take superposition of two kernels
$K^{(s)}(g_1,g_2,g_3)$ corresponding to both different irreducible
representations $s=1,2$ given by (\ref{R8}) and (\ref{R13}). This
superposition satisfies the associativity equation being
determined by the convolution product (\ref{R19}).

\subsection{Quaternionic  group}

We consider a group $G$ with eight elements in the representation provided by the Pauli
matrices. If we properly redefine the trace, they are orthonormal.
We have
\begin{eqnarray}\label{G81}
&&E=\sigma_0,\qquad P=-\sigma_0,\qquad
K=i\sigma_1,\qquad L=i\sigma_2,\nonumber\\
&&\\
&&M=i\sigma_3,\qquad K'=-i\sigma_1,\qquad L'=-i\sigma_2,\qquad M'=-i\sigma_3,\nonumber
\end{eqnarray}
where the space of functions $f:\left\{0,1,2,\ldots,7\right\}\to{\mathbf C}$ is represented
by ${\mathbf C}^{8}$ and we find the kernel function given below.

The abstract group multiplication table obtained using explicit form of the Pauli matrices
\begin{eqnarray}\label{G82}
&&\sigma_0=\left(\begin{array}{cc} 1&0\\0&1\end{array}\right),\qquad
\sigma_1=\left(\begin{array}{cc} 0&1\\1&0\end{array}\right),\nonumber\\
&&\\
&&\sigma_2=\left(\begin{array}{cc} 0&-i\\i&0\end{array}\right),\qquad
\sigma_3=\left(\begin{array}{cc} 1&0\\0&-1\end{array}\right)\nonumber
\end{eqnarray}
reads
\begin{eqnarray}\label{G83}
&&P^2=E,\quad K^2=L^2=M^2=K'^2=L'^2=M'^2=P,\quad KL=M,\nonumber\\
&&LM=K,\quad MK=L,\quad K'L'=M',\quad L'M'=K',\quad M'K'=L',\\
&&K'K=LL'=MM'=E,\quad LK=M',\quad ML=K',\quad KM=L'.
\nonumber
\end{eqnarray}
Other products of the group elements follow from this table easily.
The group has five irreducible representations due to the decomposition
$$8=1^2+1^2+1^2+1^2+2^2.$$ There are four one-dimensional and one two-dimensional
representations. The representations are unitary ones. The characters of the representations
are given in the following table of characters:
$$\begin{array}{cccccccc}
E&P&L&K&M&L'&K'&M'\\
1&1&1&1&1&1&1&1\\
1&1&1&-1&-1&1&-1&-1\\
1&1&-1&1&-1&-1&1&-1\\
1&1&-1&-1&1&-1&-1&1\\
2&-2&0&0&0&0&0&0
\end{array}$$
The first four rows in the table are characters of the one-dimensional representations
satisfying the rules of group-element multiplication given in table (\ref{G83}). The last
row contains characters of two-dimensional irreducible representation given by (\ref{G81}).

Now one can construct kernel of star product for a given group following the method described
and using explicitly the properties of the Pauli matrices.

If we denote the elements in (\ref{G81}) as
\begin{eqnarray}\label{G84}
&&E=g_1,\qquad P=g_{-1},\qquad
K=g_2,\qquad L=g_3,\nonumber\\
&&\\
&&M=g_4,\qquad K'=g_{-2},\qquad L'=g_{-3},\qquad M'=g_{-4}\nonumber
\end{eqnarray}
and apply for the two-dimensional representation the formula for characters,
we obtain the star-product-structure constants which are nonzero of the form
\begin{equation}\label{G85}
K_{mn}^s=\frac{1}{4}\left[\chi\left(g_mg_ng_s^{-1}\right)\right].\end{equation}
Here only elements $K_{mn}^{\pm 1}$, $K_{\pm 1n}^{m}$, and $K_{n\pm
1}^{m}$ with $n=\pm m$ differ from zero, the other
structure-constant elements are zero. For example,
\begin{equation}\label{G86}
K_{11}^1=-K_{11}^{-1}=K_{-11}^{-1}=-K_{1-1}^1=1/2.
\end{equation}
Since the antisymmetric part of the kernel of the star product is nonzero, the corresponding Lie
algebra structure constants read
\begin{equation}\label{G87}
C_{mn}^s=K_{mn}^s-K_{nm}^s=
\frac{1}{4}\mbox{Tr}\left(\left[g_m,g_n\right]g_s^{-1}\right).\end{equation}
The obtained Lie algebra is a subalgebra of the eight complex linear
transformations acting on ${\mathbf C}^2$ which define a
four-dimensional complex vector space. By construction, the Lie
algebra coincides with the Lie algebra of the $GL(2,C)$ group, i.e.,
the complexification of the Lie algebra of $U(2)$.

There exists another finite group with 8 elements.
This group is the symmetry group of the square on a plane with four vertices
at $(1,1)$, $(1,-1)$, $(-1,1)$, and $(-1,-1)$.
This group contains four reflections with respect to lines which are axes of
the Cartesian coordinates and the axes rotated by angle $2\pi/4$. There are also three rotations by angles
$2\pi/4$, $2\pi/2$, and $3\pi/2$ denoted, respectively, as $C_4$, $C_4^2$, and $C_4^2$ and the
identity element $E$. We denote these elements as $E$, $C_4$, $C_4^2$, $C_4^3$, $\Sigma_1$,
$\Sigma_2$, $\sigma_{13}$, and $\sigma_{24}$. The elements  $\Sigma_1$ and $\Sigma_2$ are
reflections in the ordinate and abscissa lines, respectively, and the elements
$\sigma_{13}$ and $\sigma_{24}$ are reflections with respect to bisectrices connecting vertices of the square.
The group has a two-dimensional representation realized by Pauli matrices of the form
$$
\begin{array}{cccccccc}
E&C_4&C_4^2&C_4^3&\Sigma_1&\Sigma_2&\sigma_{13}&\sigma_{24}\\
\sigma_0&i\sigma_{3}&-\sigma_{0}&-i\sigma_{3}&-\sigma_{2}&\sigma_1&-\sigma_{1}&\sigma_{2}
\end{array}
$$
We denote the elements as $E=g_1$, $C_4=g_2$, $C_4^2=g_3$, $C_4^2=g_4$, $\Sigma_1=g_5$, $\Sigma_2=g_6$,
 $\sigma_{13}=g_7$, and $\sigma_{24}=g_8$. Then the multiplication table for this group is
$$\begin{array}{ccccccccc}\quad
   &g_1&g_2&g_3&g_4&g_5&g_6&g_7&g_8\\
g_1&g_1&g_2&g_3&g_4&g_5&g_6&g_7&g_8\\
g_2&g_2&g_3&g_4&g_1&g_8&g_7&g_5&g_6\\
g_3&g_3&g_4&g_1&g_2&g_6&g_5&g_8&g_7\\
g_4&g_4&g_1&g_2&g_3&g_7&g_8&g_6&g_5\\
g_5&g_5&g_7&g_6&g_8&g_1&g_3&g_2&g_4\\
g_6&g_6&g_8&g_5&g_7&g_3&g_1&g_4&g_2\\
g_7&g_7&g_6&g_8&g_5&g_4&g_2&g_1&g_3\\
g_8&g_8&g_5&g_7&g_6&g_2&g_4&g_3&g_1
\end{array}$$
The table of characters of the unitary irreducible representations is given below
$$\begin{array}{cccccccc}
E&C_4&C_4^2&c_4^3&\Sigma_1&\Sigma_2&\sigma_{13}&\sigma_{24}\\
2&0&-2&0&0&0&0&0\\
1&1&1&1&1&1&1&1\\
1&-1&1&-1&-1&-1&1&1\\
1&-1&1&-1&1&1&-1&-1\\
1&1&1&1&-1&-1&-1&-1
\end{array}$$
One can see that tables of characters of quaternionic group and symmetry group of square are
identical. This means that star products described by the characters of irreducible representations
are also the same. It is not trivial and intuitively not obvious why two different groups provide the same structure constants.
In fact, these two finite groups of order eight can be mapped one into the other
to be considered as different realizations of the same abstract group.
The map consists of the shift of the group elements by left or right multiplication by another element of the group.

Let us describe the procedure in detail.

Given a group $G$ with $N$ elements $g_1,g_2,\ldots,g_N$. Let $g_1$
be identity element in the group $g_1=E$. Let us have table of multiplication in the given group like
$$g_kg_j=g_s.$$
With respect to given identity element $g_1$, one has inverse element
$g_1^{-1},g_2^{-1},\ldots,g_N^{-1}$.
Let us consider the set $\tilde G=\tilde g_1,\tilde g_2,\ldots\tilde g_N$ with
$\tilde g_k=g_kg_0,$ where $g_0$ is chosen as one of $N$ elements of the group $G$.
Now one can introduce a new (with respect to the initial one) multiplication table in the set $\tilde G$ using the following rules ($\star$ rules)
\begin{equation}\label{CC1}
\tilde g_k\star\tilde g_j=\left(g_kg_0\right)g_0^{-1}\left(g_{j}g_0\right)=\tilde g_s=g_sg_0.
\end{equation}
The structure constants of the new product coincide with those of
the previous one. The new rule uses idea of the so-called
$k$-product of matrices as mentioned in Sect. 3 and considered,
e.g., in \cite{OlgaJPA} (or $k$-deformed product of matrices)
where the rule of product -- row by column -- is modified by
inserting a chosen matrix $k$ when one multiplies two matrices $a$
and $b$. This means that
\begin{equation}\label{CC2}
a\cdot b\to a\cdot_kb=akb.
\end{equation}
This matrix product is associative. The new group multiplication
(\ref{CC1}) just uses the analog of the rule (\ref{CC2}) where the
element $g_0^{-1}$ plays the role of matrix $k$. This means that
in terms of the initial identity element $g_1$ and the
``deforming'' shift element $g_0$ the new identity in $\tilde G$
reads
\begin{equation}\label{CC3}
\tilde E=g_1g_0=g_0=\tilde g_1.
\end{equation}
In fact,
\begin{equation}\label{CC4}
\tilde g_k\cdot\tilde g_1=
g_kg_0g_0^{-1}g_1g_0=g_kg_0=\tilde g_k.
\end{equation}
Also $$
\tilde E\star\tilde g_k=
g_0g_0^{-1}g_kg_0=\tilde g_k.
$$
Thus with the new deformed multiplication rule one has the new
identity element and reproduces the multiplication table of the
initial group. Exactly this happens in the case of the
quaternionic group and the symmetry group of square. Nevertheless,
the realization of symmetry operations physically is quite
different in both cases. For example, the identity element for
symmetry of square means that one is doing no operation with
square. The identity element which in the deformed group
(quaternionic group) is reflection, physically differs from
`doing-no-operation'. These findings are in line with our general
considerations at the end of Sect.~{\ref{finite}}.

\subsection{Example of $C_{3v}$ group}

The group of permutations of three elements is the group of
symmetry of the equilateral triangle. The elements are:
\begin{equation}\label{V1}
g_1=1, \quad g_2=u_1, \quad g_3=u_2, \quad g_4=u_3, \quad g_5=C_3,
\quad g_6=C_3^2,
\end{equation}
here $C_3$ is a cyclic permutation and $u_1,u_2,u_3$ are
permutations, odd ones. There are three irreducible
representations with table of characters of the form
\begin{eqnarray}\label{V2}
\begin{array}{cccccc}
g_1&g_2&g_3&g_4&g_5&g_6\\
1&1&1&1&1&1\\
1&-1&-1&-1&1&1\\
2&0&0&0&-1&-1
\end{array}
\end{eqnarray}

Let us discuss one-dimensional representations.

Given a 1$\times$1 matrix $A$ which is a number. The
symbol of this operator reads
\begin{equation}\label{V4}
f_A(g)=Au(g)=\mbox{Tr}\left(Au(g)\right).
\end{equation}
In the case of identity representation, the reconstruction formula reads
\begin{equation}\label{V4a}
A=\frac{1}{6}\sum_{k=1}^6f_A(g_k)=\frac{1}{6}6A=A.
\end{equation}

Analogously reconstruction formula can be obtained for the second
one-dimensional representation with characters given in the second
line of (\ref{V2}).

The considered operators $A$ and $B$, acting in a one-dimensional Hilbert
space, are numbers. The product of
two operators $AB$ is just product of these numbers. The
star-product of the symbols in this case reads
\begin{equation}\label{V6}
f_{AB}(g)=\mbox{Tr}\left(ABu(g)\right)=ABu(g)=f_A(g)\star f_B(g).
\end{equation}

Let us check that this formula is coherent with the formula with
the star-product kernel
\begin{eqnarray}\label{V7}
f_A(g)\star
f_B(g)&=&\sum_{g_1g_2}\left[\mbox{Tr}\left(\frac{1}{N}u^{-1}(g_1)
\frac{1}{N}u^{-1}(g_2)u(g)\right)\right]f_A(g_1)f_B(g_2)\nonumber\\
&=&\frac{1}{N^2}\sum_{g_1g_2}Au(g_1)Bu(g_2)u^{-1}(g_1)u^{-1}(g_2)u(g)\nonumber\\
&=&ABu(g)\end{eqnarray} Thus we checked that the formula yields
the result shown in (\ref{V6}). Now one can apply the same kernel
to use the star-product of functions on the whole group. In this
case, a function on the group can be considered as a 6-vector. The
product of the functions is equivalent to the star-product of two
6-vectors. If one uses as a kernel of the star-product the
character of identity representation, one gets
\begin{equation}\label{R14b}
\vec f_1\star\vec f_2=\vec f,
\end{equation}
where the 6-vector $\vec f$ has all six components equal to $x$ and this
number $x$ is expressed in terms of the components of the vectors
$f_{1s}$ and $f_{2k}$ as
\begin{equation}\label{R15b}
f=\frac{1}{36}\sum_{k=1}^6\sum_{s=1}^6f_{1k}f_{2s}.
\end{equation}

In the case of antisymmetric representation, the components of the
vector $\vec f$ have different signs corresponding to even and odd
permutations. One can check that these two star-products are not
equivalent.

The Lie algebra structure constants are equal to zero if
characters of irreducible representations have the properties
\begin{equation}\label{R16b}
\chi\left(g_1g_2g_3^{-1}\right)=
\chi\left(g_2g_1g_3^{-1}\right)\quad\mbox{or}\quad
\chi\left(g_1^{-1}g_2^{-1}g_3\right)=
\chi\left(g_2^{-1}g_1^{-1}g_3\right)
\end{equation}
for all the elements $g_1$, $g_2$, $g_3$.

In the case of $C_{3v}$ group, for all its one-dimensional irreducible
representations, equality (\ref{R16b}) holds. In view of this, the
Lie algebras associated with these irreducible representations of this
group are Abelian ones.

In the case of $k$-deformed star-product, one has deformed kernels
\begin{equation}\label{R17b}
K_k(g_1,g_2,g_3)=\frac{N_s^2}{N^2}\mbox{Tr}\left(ku(g_2^{-1}g_3g_1^{-1})
\right)=\frac{N_s^2}{N^2}f_k(g_2^{-1}g_3g_1^{-1})
\end{equation}
and
\begin{equation}\label{R18b}
K_k^d(g_1,g_2,g_3)=\frac{N_s}{N}\mbox{Tr}\left(ku(g_2g_3^{-1}g_1)
\right).
\end{equation}

Let us consider in detail the two-dimensional representation of the
group consisting of 6 elements $g_k$ (or group  $C_{3v}$). Its
table of multiplication reads
\begin{eqnarray*}
&&g_1g_1=g_1,\quad g_1g_2=g_2,\quad g_1g_3=g_3,\quad
g_1g_4=g_4,\quad g_1g_5=g_5,\quad g_1g_6=g_6,\\
&&g_1g_2=g_2,\quad g_2g_2=g_3,\quad g_2g_3=g_1,\quad
g_2g_4=g_5,\quad g_2g_5=g_6,\quad g_2g_6=g_4,\\
&&g_3g_1=g_3,\quad g_3g_2=g_1,\quad g_3g_3=g_2,\quad
g_3g_4=g_6,\quad g_3g_5=g_4,\quad g_3g_6=g_5,\\
&&g_4g_1=g_4,\quad g_4g_2=g_6,\quad g_4g_3=g_5,\quad
g_4g_4=g_1,\quad g_4g_5=g_3,\quad g_4g_6=g_2,\\
&&g_5g_1=g_5,\quad g_5g_2=g_4,\quad g_5g_3=g_6,\quad
g_5g_4=g_2,\quad g_5g_5=g_1,\quad g_5g_6=g_3,\\
&&g_6g_1=g_6,\quad g_6g_2=g_5,\quad g_6g_3=g_4,\quad
g_6g_4=g_3,\quad g_6g_5=g_2,\quad g_6g_6=g_1.\end{eqnarray*}
The matrices of two-dimensional irreducible representation read
\begin{eqnarray*}
&& g_1=\left(
\begin{array}{cc}
1&0\\0&1\end{array} \right),\qquad
g_2=\left(
\begin{array}{cc}
\varphi&0\\0&\varphi^{-1}\end{array} \right),\qquad g_3=\left(
\begin{array}{cccc}
\varphi^2&0\\0&\varphi^{-2}\end{array} \right),\\
&& g_4=\left(
\begin{array}{cc}
0&1\\1&0\end{array} \right),\qquad g_5=\left(
\begin{array}{cc}
0&\varphi\\
\varphi^{-1}&0\end{array} \right),\qquad g_1=\left(
\begin{array}{cc}
0&\varphi^2\\
\varphi^{-2}&0\end{array} \right), \end{eqnarray*}
where $\varphi= e^{2\pi i/3}$ corresponds to rotation by $2\pi/3$.

According to the construction of symbol of the operator $\hat A$
with the matrix
\begin{equation}\label{W1}
 A=\left(
\begin{array}{cc}
a_{11}&a_{12}\\a_{21}&a_{22}\end{array}\right)
\end{equation}
one has the following values of the function $f(g)$ on the
permutation group
\begin{equation}\label{W1a}
 f_A(g_k)=\frac{1}{3}\mbox{Tr}\left(
\begin{array}{cc}
a_{11}&a_{12}\\a_{21}&a_{22}\end{array}\right)g_k^{-1}.
\end{equation}
The reconstruction formula for the matrix $A$ can be written in
terms of quantizer operator and it reads
\begin{eqnarray}\label{W3}
A&=&\frac{1}{3}\left[\left(a_{11}+a_{22}\right)g_1+\left(a_{11}\varphi
+a_{22}\varphi^{-1}\right)g_3+\left(a_{11}\varphi^{-1}+a_{22}\varphi\right)g_2\right.\nonumber\\
&&\left.\left(a_{12}+a_{21}\right)g_4+\left(a_{12}\varphi^{-1}
+a_{21}\varphi\right)g_5+\left(a_{12}\varphi+a_{21}\varphi^{-1}\right)g_2\right].\end{eqnarray}
We used dual formula for star-product. The Lie algebra constructed
by means of the structure constants obtained with the character
formula yields in the basis
\begin{eqnarray}\label{W4}
&&y_1=g_2-g_3,\qquad y_2=g_4,\qquad y_3=g_5,\nonumber\\
&&\\
&&y_4=g_1,\qquad y_5=g_2+g_3,\qquad y_6=g_4+g_5+g_6\nonumber
\end{eqnarray}
the following commutation relations:
\begin{eqnarray}\label{W5}
&&[y_1,y_2]=2y_2+4y_3-2y_6,\qquad [y_2,y_3]=-y_5,\nonumber\\
&&\\
&&[y_3,y_1]=2y_3+4y_2-2y_6,\qquad [y_1,y_6]=[y_2,y_6]=[y_3,y_6]=0.
\nonumber
\end{eqnarray}
The operators $y_4$, $y_5$, and $y_6$ commute with all the other
operators. The Lie algebra obtained is a direct sum of a
five-dimensional Lie algebra and a one-dimensional one. The
five-dimensional Lie algebra is an extension of the
Heisenberg--Weyl algebra defined by $(y_2,y_3,y_5)$. The
nontrivial structure constants are
\begin{equation}\label{W6}
 C_{12}^2=2,\quad C_{12}^3=4,\quad
 C_{23}^5=-1,\quad C_{31}^2=4,\quad
 C_{31}^3=2,\quad C_{12}^6=-2,\quad C_{31}^6=-2.
\end{equation}

Analogously, $k$-deformed kernel can be constructed for star product.
Also $k$-deformed Lie-algebra structure constants can be expressed
in terms of the characters.

\subsection{Example of $SU(2)$ group}

The construction can be applied also for a compact group $G$. One
has only to make change in (\ref{10}) since instead of sum one has integral
over compact group, i.e.,
\begin{equation}\label{K1}
\int
d\mu(g)\,u^{(s)}(g)_{mn}u^{\nu(p)}_{\alpha\beta}(g)=\delta_{m\alpha}
\delta_{u\beta}\frac{V}{N_s}\delta_{sp},
\end{equation}
where $d\mu(g)$ is invariant Haar measure and
\begin{equation}\label{K2}
\int d\mu(g)=V
\end{equation}
is group volume. Superindices $(s)$ and $(p)$ describe the Casimir
operators eigenvalues distinguishing different irreducible
representations of the compact groups.

For spin $j=1/2$ (defining representation), one has that
\begin{equation}\label{K3}
 g=\left(
\begin{array}{cc}
\alpha &\beta\\
-\beta^*&\alpha^*\end{array}\right),\quad
\alpha=\cos\frac{\theta}{2}e^{i(\varphi+\psi)/2},\quad
\beta=\sin\frac{\theta}{2}e^{i(\varphi-\psi)/2}.
\end{equation}
The symbol of a spin operator (matrix $A$) reads
\begin{equation}\label{K4}
 f_A(g)=\mbox{Tr}\left(
\begin{array}{cc}
A_{11}&A_{12}\\
A_{21}&A_{22}\end{array}\right)\left(
\begin{array}{cc}
\alpha &\beta\\
-\beta^*&\alpha^*\end{array}\right).
\end{equation}
One has inverse relation using the quantizer
\begin{equation}\label{K5}
A= \int\frac{2}{V} d\mu(g)\,f_A(g)g^{-1}.
\end{equation}
The kernel of star-product reads
\begin{equation}\label{K6}
 K(g_1,g_2,g_3)=\frac{4}{V^2}\mbox{Tr}\left[
 \left(
\begin{array}{cc}
\alpha_3^* &-\beta_3\\
\beta_3^*&\alpha_3\end{array}\right)
 \left(
\begin{array}{cc}
\alpha_2^*&-\beta_2\\
\beta_2^*&\alpha_2\end{array}\right) \left(
\begin{array}{cc}
\alpha_1 &\beta_1\\
-\beta_1^*&\alpha^*_1\end{array}\right)\right].
\end{equation}
The dual kernel reads
\begin{equation}\label{K7}
 K^{(d)}(g_1,g_2,g_3)=\frac{2}{V}\mbox{Tr}\left[
 \left(
\begin{array}{cc}
\alpha_1&\beta_1\\
-\beta_1^*&\alpha^*_1\end{array}\right)
 \left(
\begin{array}{cc}
\alpha_2&\beta_2\\
-\beta_2^*&\alpha_2^*\end{array}\right) \left(
\begin{array}{cc}
\alpha_3^* &-\beta_3\\
\beta_3^*&\alpha_3\end{array}\right)\right].
\end{equation}
Analogous explicit formulae can be given using known matrix
elements of irreducible representations of $SU(2)$ group in terms
of Euler angles. The star-product of two functions on $SU(2)$
group given as the kernel (\ref{K6}) reads
\begin{equation}\label{K7a}
\left(f_1\star f_2\right)(g)=\int K(g_1,g_2,g)f_1(g_1)f_2(g_2)\,d\mu(g_1)
\,d\mu(g_2),
\end{equation}
where the $SU(2)$-group elements $g_1$, $g_2$, $g$ are labelled by
Euler angles. The measures $d\mu(g_1)$ and $d\mu(g_2)$ are known
Haar measures of the $SU(2)$ group. The kernel can be taken from
(\ref{K6}) and (\ref{K7}).

One can easily calculate the Lie algebra structure constants by calculating traces
\begin{equation}\label{K6a}
 C(g_1,g_2,g_3)=\frac{4}{V^2}\mbox{Tr}\left[\left[
 \left(
\begin{array}{cc}
\alpha_3^* &-\beta_3\\
\beta_3^*&\alpha_3\end{array}\right),
 \left(
\begin{array}{cc}
\alpha_2^*&-\beta_2\\
\beta_2^*&\alpha_2\end{array}\right)\right] \left(
\begin{array}{cc}
\alpha_1 &\beta_1\\
-\beta_1^*&\alpha^*_1\end{array}\right)\right]
\end{equation}
and
\begin{equation}\label{K7a}
 C^{(d)}(g_1,g_2,g_3)=\frac{2}{V}\mbox{Tr}\left[\left[
 \left(
\begin{array}{cc}
\alpha_1&\beta_1\\
-\beta_1^*&\alpha^*_1\end{array}\right),
 \left(
\begin{array}{cc}
\alpha_2&\beta_2\\
-\beta_2^*&\alpha_2^*\end{array}\right)\right] \left(
\begin{array}{cc}
\alpha_3^* &-\beta_3\\
\beta_3^*&\alpha_3\end{array}\right)\right].
\end{equation}
The structure constant are not zero.

In fact, we present Lie group structure constants (\ref{K7a}) in explicit form
\begin{eqnarray}\label{last}
&&C^{(d)}\left(\alpha_1,\beta_1,\alpha_2,\beta_2,\alpha_3,\beta_3\right)=\frac{2}{V}
\left\{\left(\beta_2\beta_1^*-\beta_1\beta_2^*\right)\alpha_3^*+
\left(\alpha_1\beta_2-\alpha_2\beta_1+\beta_1\alpha_2^*-\beta_2\alpha_1^*\right)
\beta_3^*\right.\nonumber\\
&&\left.-\beta_3\left(\beta_2^*\alpha_1-\alpha_2^*\beta_1^*+\alpha_1^*\beta_2^*-\beta_1^*\alpha_2
\right)+\alpha_3\left(\beta_2^*\beta_1-\beta_1^*\beta_2\right)\right\}.
\end{eqnarray}

The structure constants both for star product and Lie product are strongly related
to the used irreducible two-dimensional representation of $SU(2)$ group. If one considers
an arbitrary function on $SU(2)$ group $\Phi(\alpha,\beta)$, it can be decomposed into
series connecting all the irreducible representations. The star-product kernel constructed
makes projection to the components in these series which belong to the chosen irreducible
representation. One can see this phenomenon for the group of two elements.

Thus, the functions (vectors) of the form
$$\mid f\rangle=\left(\begin{array}{c}f\\f\end{array}\right)$$
being multiplied by kernel induced by identity representation keep this form yielding
result (\ref{R18a}).

The functions (vectors) of the form
$$\mid\Phi\rangle=\left(\begin{array}{c}x\\-x\end{array}\right)$$
being multiplied by the same kernel yield as result zero function
$$\mid\Phi_1\rangle\star\mid\Phi_2\rangle=0.
$$

Thus the kernel provides projection on the functions corresponding to the
irreducible representation.


\section{Conclusions} \label{conclusions}

We now point out the main results of the paper.

We have shown that there exists a star product of complex-valued
functions on finite and compact groups. The kernels generating
such products are expressed in terms of characters of irreducible
unitary representations of these groups. Thus, the known tables of
the characters induce star products on the functions over the
groups (in group algebras). The relations of the kernels
associated with different irreducible representations (for
example, the compatibility of the structure constants) needs
further study. The star-product kernels provide in the standard
manner Lie algebra structure constants. Therefore, we found a
relation between finite (and compact) group irreducible
representations and star-product kernels along with the structure
constants of the related Lie algebras. We hope to study in detail
the mutual compatibility of the obtained structure constants in a
future publication. In particular, we shall address the
decomposition of the group algebra into minimal ideals and the
structure of the space of unitary representations of a given
finite group $G$.

\section*{Acknowledgements}

V.I.M. thanks Depto.\ de Matem\'aticas, Univ.\ Carlos III de
Madrid for kind hospitality and the Russian Foundation for Basic
Research for partial support under Projects Nos.~07-02-00598 and
08-02-90300.

\end{document}